\documentclass[a4,11pt]{article}

\usepackage{amsmath, amssymb}

\usepackage{epsfig}
\usepackage{placeins}
\usepackage{array}
\usepackage{amsmath}
\usepackage[utf8]{inputenc}
\usepackage{multirow}
\usepackage[table]{xcolor}
\usepackage{mdwlist}
\usepackage{natbib}
\usepackage{dsfont}

\newtheorem{rmk}{\sc Remark}

\def \si {\sigma}

\def \beg{\begin{eqnarray}}
\def \en{\end{eqnarray}}
\def \be*{\begin{eqnarray*}}
\def\e*{\end{eqnarray*}}
\def \di{\displaystyle}
\def\bit{\begin{itemize}}
\def \eit{\end{itemize}}
\def \E{\mathbb E}

\begin{document}

\begin{center}
{\Large
	{\sc
	 A study of variable selection using $g$-prior distribution with ridge parameter
	}
}
\bigskip

Me\"{i}li Baragatti$^{1,2,*}$ and Denys Pommeret$^2$

\medskip
{\it
 $^1$ Ipsogen SA, Luminy Biotech Entreprises, Case 923, Campus de Luminy, 13288 Marseille Cedex 9, France.\\
 $^2$ Institut de Mathématiques de Luminy (IML), CNRS Marseille, case 907, Campus de Luminy, 13288 Marseille Cedex 9, France.\\
 $^*$ baragatt@iml.univ-mrs.fr, baragattimeili@hotmail.com.
}
\end{center}

\begin{center}
{\sc Working Paper \today}
\end{center}

\bigskip
\noindent

	\begin{abstract}
	In the Bayesian stochastic search variable selection framework, a common prior distribution for the regression coefficients is the $g$-prior of \citet{Zellner86}. However, there are two standard cases in which the associated covariance matrix does not exist, and the conventional prior of Zellner can not be used:
	if the number of observations is lower than the number of variables (large $p$ and small $n$ paradigm), or if some variables are linear combinations of others.
	In such situations a prior distribution derived from the prior of Zellner can be used, by introducing a ridge parameter. This prior introduced by  \cite{GuptaIbrahim} is a flexible and simple adaptation of the $g$-prior.
	In this paper we study the influence of the ridge parameter on the selection of variables.  A simple way to choose the associated hyper-parameters is proposed. The method is valid  for any generalized linear mixed  model and we focus on the case of  probit mixed models when some variables are linear combinations of others. The method is  applied to both simulated and real datasets obtained from Affymetrix microarray experiments. Results are compared to those obtained with the Bayesian Lasso.
	\end{abstract}

\section{Introduction}\label{intro}

	We consider the problem of Bayesian variable  selection in a generalized linear mixed model with $Y$ a $n$-vector of responses,
	given a set of $p$ potential fixed regressors
	\begin{displaymath}
		g(\E(Y_i  \mid  U, \beta))=X_i^T\beta + Z_i^TU,
	\end{displaymath}
	where $g$ stands for the link function associated to the model,
	and $X_i$ and $Z_i$ for the fixed and random effect regressors associated to the $i${th} observation. The parameter $\beta\in\mathbb R^p$ corresponds to the fixed-effect coefficients and the parameter $U$ to the random-effect coefficients. $X$ and $Z$ are known design matrices associated with the fixed and random effects. We consider $K$ random effects, $U=(U_1^T, \cdots, U_K^T)^T$ where each $U_l$ is a vector of size $q_l$, and $\sum_{l=1}^K q_l = q$.

	In a stochastic search variable selection (SSVS) framework, it is convenient to denote by
	$\gamma$ the vector of latent variables indicating if a variable is selected or not; that is,
	$\gamma_j=1$ if $\beta_j \neq 0$ and $\gamma_j=0$ if $\beta_j =0$.
	We then denote by $\beta_{\gamma}$ the vector of all non-zero elements of $\beta$ and by $\mathbf{X}_{\gamma}$ the design matrix
	with columns corresponding to the elements of $\gamma$ that are equal to 1.\\

	To complete the model, a conventional prior distribution for $\beta_{\gamma}|\gamma$ is a $d_{\gamma}$-dimensional Gaussian distribution, with $d_{\gamma}=\sum_{j=1}^p \gamma_j$,
	\begin{equation}\label{priorbeta}
	\beta_{\gamma} |\gamma \sim {\cal N}_{d_{\gamma}}(0, \Sigma_{\gamma}).
	\end{equation}
	Concerning the prior covariance matrix $ \Sigma_{\gamma}$, an attractive and standard choice is
	\begin{equation}\label{sigmagamma}
	\Sigma_{\gamma}  =  \tau (\mathbf{X}_{\gamma}' \mathbf{X}_{\gamma})^{-1}.
	\end{equation}
	Equations (\ref{priorbeta}) and (\ref{sigmagamma})  correspond to the {\it g}-prior distribution, proposed by \cite{Zellner86} in the case of standard linear models. This prior replicates the covariance structure of the design and enables an automatic scaling based on the data. Up to the scalar $\tau$, the prior covariance matrix is related to the Fisher Information Matrix in the linear model  \citep[see for instance][]{ChenIbrahim}. Moreover, it leads to simple expressions of the marginal likelihood, and as pointed out in \cite{GeorgeFoster}, the marginal likelihood becomes a function of both R-square and the number of covariates like in AIC or  BIC criteria. The parameter $\tau>0$ is referred to as the variable selection coefficient in \cite{BottoloRichardson}. In the homoscedastic linear model with variance $\sigma^2$, this parameter can be expressed as  $\tau = g \si^2$.
	Therefore this prior has been used by many authors in the case of linear models, but also for generalized linear models (see \cite{SabanesHeld}).  In case of a binary response variable, this prior is frequently encountered in probit models which are quite practical in a Bayesian setting  \citep[see][]{LeeSha,ShaVannucci,ZhouWang1,YangSong}.

	The choice of the variable selection coefficient $\tau$ can have a great influence on the variable selection process   \citep[see][]{GeorgeFoster} and has been considered by many authors. Some of them considered a fixed value for $\tau$.
 	For instance \cite{SmithKohn}  suggested to choose $\tau$ between 10 and 100. Another approach is the approach of \cite{GeorgeFoster}, who developed empirical Bayes  methods based on the estimation  of $\tau$ from its marginal likelihood. Other authors proposed to put a hyper-prior distribution on $\tau$, like \cite{ZellnerSiow} that  used an inverse-gamma distribution $\mathcal{IG}(1/2,n/2)$. But under the Zellner-Siow prior, marginal likelihoods are not available in closed forms, and approximations are necessary  \citep[see][]{BottoloRichardson}.

	Note also that the  Zellner-Siow prior can be seen as a mixture of \textit{g}-priors. Following this remark, \cite{LiangPaulo} proposed a new family of priors on $\tau$, the \textit{hyper-g prior family} which leads to new mixtures of $g$-priors: the marginal likelihoods are then in closed forms, but not in a practical way because hypergeometric functions are used. Independently but in the same spirit, \cite{CuiGeorge} suggested to put an inverse-gamma prior distribution on $(1+\tau)$ (rather than on $\tau$ like Zellner and Siow), obtaining a family of priors on $\tau$ which contains the \textit{hyper-g prior family} as a special case. \cite{BottoloRichardson} used a similar prior. In the linear regression framework, \cite{CeleuxMarinRobert} and \cite{MarinRobert2007} suggested an improper discrete prior on $\tau$. But this prior is difficult to use in practice because it induces an infinite sum. As a consequence, \cite{CeleuxAnbari2011} proposed a Jeffrey prior continuous on $\tau$, and \cite{GuoSpeckman} showed the consistence of associated Bayes factors.


	In spite of the variety of all these works to choose the variable selection coefficient $\tau$, a crucial problem remains with priors using the matrix $(\mathbf{X}_{\gamma}^T \mathbf{X}_{\gamma})^{-1}$. Indeed, $\mathbf{X}_{\gamma}^T \mathbf{X}_{\gamma}$ should be invertible. However, there are two standard cases where  $\mathbf{X}_{\gamma}^T \mathbf{X}_{\gamma}$ is singular:
	\begin{itemize}
	\item If the number of observations is lower than the number of variables in the model, $n < d_{\gamma}$.
	\item If some variables are linear combinations of others. In practice, even if $\mathbf{X}_{\gamma}^T \mathbf{X}_{\gamma}$ is theoretically invertible, some variables can be highly correlated and $\mathbf{X}_{\gamma}^T \mathbf{X}_{\gamma}$ can be computationally singular. It is often the case in genomic high-dimensional datasets for example.  This problem can also be encountered when several datasets are merged: some variables can be collinear or almost collinear if same variables were present into several datasets under different labels for instance.
	\end{itemize}

	In these cases the classical $g$-prior does not work. Concerning the first case, several authors proposed alternative priors.
	\cite{MaruyamaGeorge} proposed a generalization of the $g$-prior, working with a singular value decomposition of the design matrix $X$. But their approach is valid only in case of classical linear models. \cite{YangSong} proposed to replace the matrix $(\mathbf{X}_{\gamma}^T \mathbf{X}_{\gamma})^{-1}$ in $ \Sigma_{\gamma}$ by its Moore Penrose's inverse (see also\cite{West}). However, the computation of the posterior distribution has a technical issue that do not permit the use of MCMC algorithm (see \cite{CommentYangSong}).
	Another idea would be to avoid this first case by fixing the number of selected covariates at each iteration, as in \cite{Baragatti1}.
	It appeared computationally advantageous and it reduced the effect of the variable selection coefficient $\tau$ used in the $g$-prior.
	But the number of selected variables at each iteration must be arbitrarily fixed. 
	Moreover, fixing the number of selected covariates is not a solution for the second case, as well as the priors proposed by \cite{MaruyamaGeorge} and \cite{YangSong}.
	In a spirit of ridge regression (see \cite{Marquardt}), \cite{GuptaIbrahim} proposed an extension of the $g$-prior, by introducing a ridge parameter. Their prior can be used in the two cases, but they did not study the second case in which some variables are linear combinations of others. Recently \cite{KwonVannucci2011} proposed a variable selection method which take into account high correlations between predictors, but again their approach is not valid when some variables are linear combinations of others. More generally, to our knowledge the problem of variable selection when some variables are linear combinations of others is not present in literature.

	In this paper we develop the idea of \cite{GuptaIbrahim} concerning the introduction of a ridge parameter, and we study the influence of this parameter on the selection of variables. Besides, we suggest a way to choose the associated hyper-parameters: following the original idea of \cite{Zellner86} which is to keep the covariance structure of the design, we propose to keep the total variance of the data through the trace of $\mathbf{X}^T \mathbf{X}$. We focused on probit models, as studied in \cite{Baragatti1}, \cite{YangSong} and \cite{LeeSha}. The aim is to study the behavior of the variable selection process while using the proposed prior, especially when some variables are linear combinations of others. The approach developed is applied both to simulation data, and to data obtained from Affymetrix microarray experiments. We compare the numerical results with those obtained by a Bayesian Lasso approach (see \cite{ParkCasella2008} and \cite{Hans2009} for a recent review) in the context of probit mixed models.

	This paper is organized as follows. In Section \ref{Ridge} the extension of the prior to be used for $\beta_{\gamma}$ is introduced and a choice for the hyper-parameters is suggested. The Section \ref{probit} outlines the priors, full conditional distributions and the sampler to be used in case of a probit mixed model, for both the SSVS approach and a Bayesian Lasso approach. In Section \ref{Results} experimental results are given and a sensitivity analysis is performed. Finally Section \ref{Discussion} discusses the method.

\section{Introducing a ridge parameter}\label{Ridge}

	\subsection{Prior distribution of $\beta$ with a ridge parameter}\label{partpriorbeta}
		As previously explained, in the case of singularity of the matrix $\mathbf{X}_{\gamma}^T \mathbf{X}_{\gamma}$, the classical $g$-prior can not be used. \cite{GuptaIbrahim} proposed to use a ridge parameter, denoted $\lambda>0$, by replacing in (\ref{sigmagamma}) the matrix  $\tau^{-1}  \mathbf{X}_{\gamma}^T\mathbf{X}_{\gamma}$ by $\tau^{-1}\big(   \mathbf{X}_{\gamma}^T\mathbf{X}_{\gamma}+ \lambda I\big)$. Imitating \cite{GuptaIbrahim} we write
			\begin{equation}\label{sigmagammaridge}
		\Sigma_{\gamma}(\lambda) = (\tau^{-1} \mathbf{X}_{\gamma}^T\mathbf{X}_{\gamma}+\lambda I)^{-1},
		\end{equation}
		and we consider the prior
		\begin{equation}\label{priorbetaridge}
		\beta_{\gamma} |\gamma \sim {\cal N}_{d_{\gamma}}\big(0, (\tau^{-1} \mathbf{X}_{\gamma}^T\mathbf{X}_{\gamma}+\lambda I)^{-1}\big) \qquad \textrm{with} \qquad d_{\gamma}=\sum_{j=1}^p \gamma_j.
		\end{equation}
		Since $\lambda$ is strictly positive, the matrix $\Sigma_{\gamma}(\lambda)$ is always of full rank and (\ref{priorbetaridge}) can be viewed as a modified form of the $g$-prior, which is a compromise between independence and instability. Indeed, for large values of $\lambda$ and $\tau$, $\Sigma_{\gamma}(\lambda)$ is close to a diagonal matrix that coincides with the conditional independent case.
		On the opposite, for small values of $\lambda$ and  $\tau$, the term $\tau^{-1} \mathbf{X}_{\gamma}^T\mathbf{X}_{\gamma}$ prevails and the inverse of $\tau^{-1} \mathbf{X}_{\gamma}^T\mathbf{X}_{\gamma}+\lambda I$ will be instable if $\mathbf{X}_{\gamma}^T\mathbf{X}_{\gamma}$ is singular. In that case, the prior distribution (\ref{priorbetaridge}) is close to the  $g$-prior case.

	\subsection{Calibrating hyper-parameters}\label{calibrate}
		Following \cite{Zellner86}, our purpose is to use the design to calibrate the covariance of $\beta_{\gamma}$ with a ridge parameter.
		Write $\Sigma_{\gamma}(0)=\tau_0 (\mathbf{X}_{\gamma}^T\mathbf{X}_{\gamma})^{-1}$, with $\tau_0$ the fixed hyper-parameter used in this classical prior. Using $\Sigma_{\gamma}(\lambda)$ instead of $\Sigma_{\gamma}(0)$ amounts to introducing a perturbation in the classical $g$-prior. An interesting feature of the classical $g$-prior is that the variance-covariance structure of the data is preserved. The ridge parameter prevents us to strictly preserve this structure. However, it is possible to replicate the total variance of the data, which corresponds, up to a normalization, to the trace of $\Sigma_{\gamma}(0)^{-1}$.
		The constraint used is then
		\begin{displaymath}
			 tr\Big(\Sigma_{\gamma}(0)^{-1}\Big)=tr\Big(\Sigma_{\gamma}(\lambda)^{-1}\Big),
		\end{displaymath}
		which yields
		\begin{equation*}
		\tau  =   \tau_0\Big[1+\di\frac{\lambda p \tau_0}{tr(\mathbf{X}_{\gamma}^T\mathbf{X}_{\gamma}) - \lambda p\tau_0}\Big],
		\end{equation*}
with the condition $\lambda p\tau_0 \neq 	tr(\mathbf{X}_{\gamma}^T\mathbf{X}_{\gamma})$. 	Concerning the choice of $\lambda$, in order to take into account the number $p$ of covariates and to reduce the effect of the ridge factor, we suggest to take $\lambda=1/p$, getting
		\begin{equation*}
		\tau  =   \tau_0\Big[1+\di\frac{\tau_0}{tr(\mathbf{X}_{\gamma}^T\mathbf{X}_{\gamma}) - \tau_0}\Big],
		\end{equation*}
with the condition $\tau_0 \neq 	tr(\mathbf{X}_{\gamma}^T\mathbf{X}_{\gamma})$.
		The vector $\gamma$ can be different between two iterations of the algorithm. Therefore we propose to use the complete design matrix $\mathbf{X}$ instead of $\mathbf{X}_{\gamma}$, yielding
		\begin{equation}\label{choixtau}
		\tau  =   \tau_0\Big[1+\di\frac{\tau_0}{tr(\mathbf{X}^T\mathbf{X}) - \tau_0}\Big],
		\end{equation}
with $\tau_0 \neq 	tr(\mathbf{X}^T\mathbf{X})$.
		In practice, the user has to choose only the parameter $\tau_0$, as $\lambda$ and $\tau$ are then obtained by $1/p$ and (\ref{choixtau}). Following \cite{SmithKohn}, $\tau_0$ could be chosen between 10 and 100, and not too close to $tr(\mathbf{X}^T\mathbf{X})$.
		It is of interest to study the influence of the hyper-parameters $\lambda$ and $\tau$. In Section \ref{sensitivity}, it will be show that these hyper-parameters do not have a large bearing on the results.

		\begin{rmk}
		The choice $\lambda=1/p$  has the advantage to be automatic and to reduce the influence of the ridge parameter when the number of variables is large. However it can lead to computational instability if this number is too large and hence $\lambda$ too small, since $\Sigma_{\gamma}(\lambda)$ is then almost singular. In our numerical study we did not encountered  this problem for $p$ around $300$. But for very large $p$   we could add a threshold $\epsilon$  and then choose $\lambda=\max(1/p, \epsilon)$.
		\end{rmk}

\section{Illustration trough a mixed probit model}\label{probit}
	\subsection{The probit mixed model}

	We consider the problem of variable selection among a set of $p$ potential fixed regressors, in the following probit mixed model
	\begin{displaymath}
		P(Y_i=1  \mid  U, \beta)=p_i=\Phi(X_i^T\beta + Z_i^TU),
	\end{displaymath}
	where $\Phi$ stands for the standard Gaussian cumulative distribution function.
	Following \cite{AlbertChib} and \cite{LeeSha}, a vector of latent variables $L=(L_1,\ldots,L_n)^T$ is introduced, and we assume that the conditional distribution of $L$ is Gaussian, that is $L  \mid  U, \beta \sim \mathcal{N}_n(X\beta+ZU, I_n)$, with $I_n$ the identity matrix. We then have
	\begin{equation}\label{LatentVar}
		Y_i = \left\{
		\begin{array}{rl}
			1 & \text{if } L_i>0\\
			0 & \text{if } L_i<0.
		\end{array} \right.
	\end{equation}
	
	\subsection{Stochastic Search Variable Selection}\label{SSVS}
		\paragraph{Prior and full conditional distributions}~~\\
		We used the following prior distributions, which are classical except the one for $\beta_{\gamma}$: 
		\begin{itemize}
		 \item As explained in Section \ref{Ridge}, we use the prior (\ref{priorbetaridge}) for $\beta_{\gamma}$.
		 \item The $\gamma_j$ are assumed to be independent Bernoulli variables, with
		\begin{equation}\label{priorgamma}
		P(\gamma_j=1) =\pi, \qquad 0 \leq \pi \leq 1,
		\end{equation}
		as we do not want to use prior knowledge to favor any variables.
		 \item The vector of coefficients associated with the random effects is assumed to be Gaussian and
		centered, with covariance matrix $D$:
		\begin{equation}\label{priorU}
		U|D \sim  {\cal N}_q(0,D).
		\end{equation}
		We will consider the case where $D$ is a diagonal matrix  $D=diag(A_1,\ldots,A_K)$, where $A_l=\sigma_l^2 I_{q_l}$, $l=1,\ldots,K$ and $I_{q_l}$ the identity matrix. The prior distributions for the $\sigma_l^2$ are then Inverse Gamma $\mathcal{IG}amma(a,b)$ ($b$ denoting the scale parameter).
		In a more general case, if no structure is assumed for the variance-covariance matrix $D$, its prior distribution should be an Inverse-Wishart.
		\end{itemize}

		\noindent Most of the full conditional distributions did not depend on the ridge parameter. In particular:
		\begin{itemize}
		\item The full conditional distribution of $L$ is given by (see \cite{AlbertChib}):
			\begin{eqnarray}\label{fullL}
			L_i|\beta, U, Y_i=1 & \sim & \mathcal{N}(X_i^T\beta+Z_i^TU,1) {\rm \ left \ truncated \ at \ } 0
			\\
			\nonumber L_i|\beta, U, Y_i=0 & \sim & \mathcal{N}(X_i^T\beta+Z_i^TU,1) {\rm \ right \ truncated \ at \ } 0.
			\end{eqnarray}

		\item Defining $W=(Z^TZ+D^{-1})^{-1}$, the full conditional distribution of $U$ is:
			\begin{equation}\label{fullU}
			U|L,\beta, D  \sim  \mathcal{N}_q(WZ^T(L-\mathbf{X}\beta),W).
			\end{equation}

		\item The full conditional distribution of the $\sigma_l^2,l=1,\ldots,K$ are Inverse-Gamma:
			\begin{eqnarray}\label{fullsigma}
				\sigma_l^2 \mid U_l  &\sim& \mathcal{IG}amma\Big(\frac{q_l}{2}+a,\big(\frac 12 U_l^TU_l+b\big)\Big).
			\end{eqnarray}
  		\end{itemize}

		Only the full conditional distributions of $\beta_{\gamma}$ and $\gamma$ depend on $\lambda$, as follows:
		\begin{itemize}
		 \item For $\beta_{\gamma}$:			
			\begin{equation}\label{fullbeta}
			\beta_{\gamma}|L,U,\gamma   \sim  \mathcal{N}_{d_{\gamma}}(V_{\gamma}\mathbf{X}_{\gamma}^T(L-ZU),V_{\gamma}), \quad \textrm{with} \quad V_{\gamma} = \Big[\frac{(1+\tau)}{\tau}\mathbf{X}_{\gamma}^T\mathbf{X}_{\gamma} + \lambda I\Big]^{-1}.
			\end{equation}
		 \item And for $\gamma$:
		\begin{eqnarray}\label{fullgamma}
		\nonumber f(\gamma |L,U,\beta_{\gamma}) & \propto &
		\frac{(2\pi)^{-\frac{d_{\gamma}}{2}}}{|\Sigma_{\gamma}(\lambda)|^{1/2}} \exp\Big[-\frac{1}{2}\big(\beta_{\gamma}^T V_{\gamma}^{-1}\beta_{\gamma} - (L-ZU)^T \mathbf{X}_{\gamma} \beta_{\gamma} - \beta_{\gamma}^T \mathbf{X}_{\gamma}^T (L-ZU) \big)\Big]\\
		& \times & \prod_{j=1}^p\pi_j^{\gamma_j}(1-\pi_j)^{1-\gamma_j}.
		\end{eqnarray}	
		\end{itemize}	

		\paragraph{The sampler}~~\\
		The posterior distribution of $\gamma$ is of particular interest for the variable selection problem. An idea is to use a Gibbs sampler to explore the full posterior distribution and to search for high probability $\gamma$ values. Simulations from all the full conditional distributions can be easily obtained, except for $\gamma$ which full conditional distribution does not correspond to a standard multivariate one. The $\gamma$ vector can be simulated either element by element, or by using a Metropolis-Hastings algorithm. In general, in the case of a high number of variables, the Metropolis-Hastings algorithm is computationally advantageous. Moreover, using a Metropolis-Hastings step in a Gibbs sampler improves the sampler in terms of variance, see \cite{MonteCarloStatMethods}. As a consequence, we decided to use a Metropolis-within-Gibbs algorithm. But even with a Metropolis-Hastings algorithm, the full conditional distribution of $\gamma$ cannot be directly simulated, since it depends on the actual value of $\beta_{\gamma}$. Following \cite{LeeSha} we then used the grouping technique of \cite{Liu}, by considering the parameters $\gamma$ and $\beta_\gamma$ jointly. The advantage of this technique is that the convergence of the Markov chain is improved, and autocorrelations are reduced, see \cite{Liu} and \cite{vanDyk}. Using this technique is equivalent to integrate the full conditional distribution of $\gamma$ in $\beta_{\gamma}$ (see \cite{Baragatti1} for more details). We then obtain:
		\begin{eqnarray}\label{fullgammaint}
		\nonumber f(\gamma |L,U) & \propto &
		\di\frac{|V_{\gamma}|^{1/2}}{|\Sigma_{\gamma}(\lambda)|^{1/2}}\exp\Big[-\frac{1}{2}(L-ZU)^T
		(I-\mathbf{X}_{\gamma}V_{\gamma}\mathbf{X}_{\gamma}^T)(L-ZU)\Big]
		\\
		& \times & \di\prod_{j=1}^p\pi_j^{\gamma_j}(1-\pi_j)^{1-\gamma_j}.
		\end{eqnarray}		
		Note that setting $\lambda=0$, we can recover the formula  corresponding to the  classical $g$-prior.

		\begin{rmk} The influence of $\tau$ appears here through the ratio $R^{1/2}=\left(\frac{|V_{\gamma}|}{|\Sigma_{\gamma}|}\right)^{1/2}$.
		We can see that
		$$
		\left\{
		\begin{array}{ll} {\rm if \ } \tau \rightarrow \infty, &  R \rightarrow |  \frac{1}{\lambda}\mathbf{X}_{\gamma}^T\mathbf{X}_{\gamma}+ I|^{-1},
		\\
		{\rm if \ } \tau \rightarrow 0,  & R \rightarrow 1,
 \\{\rm if \ } \lambda \rightarrow \infty, &  R \rightarrow 1,
		\\
		{\rm if \ } \lambda \rightarrow 0,  & R \rightarrow (\di\frac{1}{1+\tau})^{d_{\gamma}/2}.
		\end{array}
		\right.
		$$
		\end{rmk}

		The Metropolis-Hastings algorithm used to generate the $\gamma$ vector can be summarized as follows: at iteration $(i+1)$  a candidate $\gamma^*$ is proposed from $\gamma^{(i)}$, and using a symmetric transition kernel the acceptance rate is
		\begin{equation*}
		\rho(\gamma^{(i)},\gamma^*)  =
		\min \Bigg\{1,\di\frac{f(\gamma^*|L,U)}{f(\gamma^{(i)}|L,U)} \Bigg\},
		\end{equation*}
		with
		\begin{eqnarray}\label{acceptancerate}
		\nonumber  \di\frac{f(\gamma^*|L,U)}{f(\gamma^{(i)}|L,U)}  & = &
		\di\left(\frac{|V_{\gamma^*}\Sigma_{\gamma^{(i)}}|}{|\Sigma_{\gamma^*}V_{\gamma^{(i)}}|}\right)^{1/2}
		\exp\Big\{-\frac{1}{2}(L-ZU)^T
		 (\mathbf{X}_{\gamma^i}V_{\gamma^{(i)}}\mathbf{X}_{\gamma^{(i)}}^T-\mathbf{X}_{\gamma^*}V_{\gamma^*}\mathbf{X}_{\gamma^*}^T)(L-ZU)\Big\}
		\\
		& \times & \di\prod_{j=1}^p \left(\di\frac{\pi_j}{1-\pi_j}\right)^{\gamma_j^*-\gamma_j^{(i)}}, \qquad \textrm{if} \qquad \forall j \in \{1,\ldots,p\} \quad \pi_j=\pi.
		\end{eqnarray}

		The simplest way to have a symetric transition kernel is to propose a $\gamma^*$ which corresponds to $\gamma^{(i)}$ in which $r$ components have been randomly changed (see \cite{ChipmanGeorge} and \cite{GeorgeMcCulloch97}).

		\begin{rmk} The influence of $\tau$ appears via the ratio $Q^{1/2} = \left(\frac{|V_{\gamma^*}\Sigma_{\gamma^{(i)}}|}{|\Sigma_{\gamma^*}V_{\gamma^{(i)}}|}\right)^{1/2}$ that satisfies:
		$$
		\left\{
		\begin{array}{ll} {\rm if \ } \tau \rightarrow \infty, &  Q \rightarrow |  \mathbf{X}_{\gamma^*}^T\mathbf{X}_{\gamma^*}+\lambda I|\times |  \mathbf{X}_{\gamma^i}^T\mathbf{X}_{\gamma^i}+\lambda I|^{-1},
		\\
		{\rm if \ } \tau \rightarrow 0,  & Q \rightarrow 1,
 \\{\rm if \ } \lambda \rightarrow \infty, &  Q \rightarrow 1,
		\\
		{\rm if \ } \lambda \rightarrow 0,  & Q \rightarrow 1.
		\end{array}
		\right.
		$$

		\end{rmk}

	\paragraph{Post-processing}~~\\
		The number of iterations of the algorithm is $b+m$, where $b$ corresponds to the burn-in period and $m$ to the observations from the posterior distributions. For selection of variables, the sequence $\{\gamma^{(t)}=(\gamma_1^{(t)},\ldots,\gamma_p^{(t)}),t=b+1,\ldots,b+m\}$ is used. The most relevant variables for the regression model are those which are supported by the data and prior information. Thus they are those corresponding to the $\gamma$ components with higher posterior probabilities, and can be identified as the $\gamma$ components that are most often equal to 1. To decide which variables should be finally selected after a run, a confidence interval based on a Poisson distribution could be used. However we noticed that usually a reasonable number of relevant variables  can be isolated from the others using the number of selections.
		Therefore we suggest to use a box-plot of the number of iterations during which variables were selected. For each run the variables distinguishable from the others can be selected by fixing a threshold: if a variable has been selected during a number of iterations which is higher than this threshold, then the variable is kept in the final selection.

\subsection{Bayesian Lasso for probit mixed models}\label{lasso}
    A competing paradigm to the classical SSVS approach is the Bayesian Lasso framework (see \cite{ParkCasella2008}), which is inspired from the frequentist Lasso (\cite{LASSO}). In order to compare the two frameworks, we adapted the Bayesian Lasso to probit mixed models. The Bayesian Lasso has already been used for probit models (\cite{BaeMallick2004}), and also for mixed models (\cite{Legarra2011}). Combining these two approaches, we considered a fully Bayesian analysis with the following prior distributions:
    \begin{itemize}
     \item[$\bullet$] For each $\beta_j,j=1,\ldots,p$ we consider a Laplace prior: $\mathcal{L}aplace(0,1/\sqrt{\delta})$.
	This Laplace distribution can be expressed as a scale mixture of normal distributions with independent exponentially distributed variances, see \cite{AndrewsMallows1974}. This prior is then equivalent to $\beta_j \mid \lambda_j \sim \mathcal{N}(0,\lambda_j)$ and $\lambda_j \sim \mathcal{E}xpo(\delta/2)$. Writing $\Lambda=diag(\lambda_1,\ldots,\lambda_p)$, we have $\beta \mid \Lambda \sim \mathcal{N}_p(0,\Lambda)$.
     \item[$\bullet$] Concerning the random effects $U$ and the variance covariance matrix $D$, we use the same classical priors than in the SSVS approach.
     \item[$\bullet$] Following \cite{ParkCasella2008}, a hyperprior distribution is put on the Lasso parameter $\delta$: $\delta \sim \mathcal{G}amma(e,f)$, $f$ denoting the scale parameter. On the following experimental results, we found like \cite{YiXu2008} and \cite{LiDasFuLiWu2011} that the posteriors were not too sensitive to the hyperparameters $e$ and $f$, as long as they were small enough so that the hyperprior is sufficiently flat. In practice we used $e=f=1$, but results were similar with $e=f=10$ for instance.
    \end{itemize}
     The Bayesian Lasso estimates for the $\beta_j$ are then obtained by a Gibbs sampler using the following posterior distributions:
    \begin{itemize}
     \item[$\bullet$] For $L$, $U$ and the $\sigma^2_l,l=1,\ldots,K$, the full conditional distributions are the same than those in the SSVS method, using (\ref{fullL}), (\ref{fullU}) and (\ref{fullsigma}).
     \item[$\bullet$] For the $\beta$, the posterior is:
		\begin{equation}\label{fullbetalasso}
		\beta|L,U,\Lambda   \sim  \mathcal{N}_{p}(V_{\Lambda}\mathbf{X}^T(L-ZU),V_{\Lambda}) \quad \textrm{with} \quad V_{\Lambda} = \Big[\mathbf{X}^T\mathbf{X}+\Lambda^{-1}]^{-1}.
		\end{equation}
     \item[$\bullet$] The posterior distributions for the $1/\lambda_j,j=1,\ldots,p$ are inverse Gaussian:
		\begin{equation}\label{fulllambdalasso}
		 1/\lambda_j \mid \beta \sim \mathcal{IG}auss\Big(\frac{\sqrt{\delta}}{\beta_j},\delta\Big).
		\end{equation}
     \item[$\bullet$] The posterior for the Lasso parameter $\delta$ is a gamma distribution:
 		\begin{equation}\label{fulldeltalasso}
		 \delta \mid \Lambda \sim \mathcal{G}amma\Big(p+e,\big(\frac{\sum \lambda_j}{2} + \frac1f\big)^{-1}\Big).
		\end{equation}
    \end{itemize}

	\paragraph{Post-processing}~~\\
		From the results of the Bayesian Lasso we obtain posterior estimates for the $\beta_j$s and the $\lambda_j$s, and the variables can be selected by three different ways:
		\begin{enumerate}
		 \item One can select the variables corresponding to an absolute value $|\beta_j|$ higher than a threshold, like \cite{YiXu2008} or \cite{LiDasFuLiWu2011} for instance.
		 \item \cite{BaeMallick2004} among others considered that the variables associated to $\beta_j$s with smaller posterior variances have no effect and should be excluded from the model. Therefore, they proposed to select variables corresponding to high values of $\lambda_j$.
		 \item Finally, the results of the Lasso enable us to obtain posterior credible intervals (CI) for the $\beta_j$s. Hence we can select variables corresponding to a $\beta_j$ with a posterior CI which does not cover 0, see \cite{Kyung2010} for instance. 
		\end{enumerate}

\section{Experimental results}\label{Results}
	\subsection{Simulated data}\label{simulateddata}
		We simulated 200 binary observations and 300 variables, the observations being obtained using a probit mixed model with 5 of these variables and one random effect of length 4. Among the 300 variables, 280 were generated from a uniform on $[-5,5]$ and denoted by $V1,\ldots,V280$. Then 10 variables denoted by $V281,\ldots,V290$ were build to be collinear to the first 10 variables, with a factor 2: for instance $V282=2 \times V2$. One variable was build to be a linear combination of $V1$ and $V2$ ($V291=V1+V2$), and another was build to be a linear combination of $V3$ and $V4$ ($V292=V3-V4$). Finally, 8 variables were build to be linear combinations of variables 5 to 12 and variables 13 to 20 (for instance $V293=V5+V13$). The five variables used to generate the binary observations were the first five: $V1,V2,V3,V4$ and $V5$. The vector of coefficients associated with these variables was $\beta=(1,-1,2,-2,3)$. The first 100 observations were part of the training set, and the last 100 were part of the validation set. In the training and the validation sets, 25 observations were associated with each component of the random effect, whom vector of coefficients was $U=(-3,-2,2,3)$.
		We had only one random effect and the different components were supposed independent, hence we put $D=\sigma^2 I_4$.

	    \paragraph{SSVS approach using the prior with a ridge parameter}~~\\
		The objective was to assess the behavior of the proposed method when some variables are linear combinations of others, and to compare it to the case where no variable is linear combination of others. Therefore we performed 10 runs of the sampler using only the first 280 variables, and 10 runs using the 300 variables. In these two cases and for each run the same parameters were used: 5 variables were initially selected, one component of $\gamma$ was proposed to be changed at each iteration of the Metropolis-Hastings step, the prior of $\sigma^2$ was a $\mathcal{IG}(1,1)$, $\pi_j=5/280$ for all $j$ when 280 variables were kept, $\pi_j=5/300$ for all $j$ when 300 variables were kept, 4000 iterations were performed after a burn-in period of 1000 iterations, and each Metropolis-Hastings step consisted of 500 iterations.
		We decided to choose $\tau_0=50$, which is a standard choice, see \cite{SmithKohn} for instance. The parameters $\lambda$ and $\tau$ were then chosen as explained in \ref{calibrate}, yielding $\lambda=1/280$ and $\tau=50.01075$ when using 280 variables, and $\lambda=1/300$ and $\tau=50.00885$ when using 300 variables.

\begin{figure}[h!]
			\begin{center}
			\includegraphics[scale=0.45]{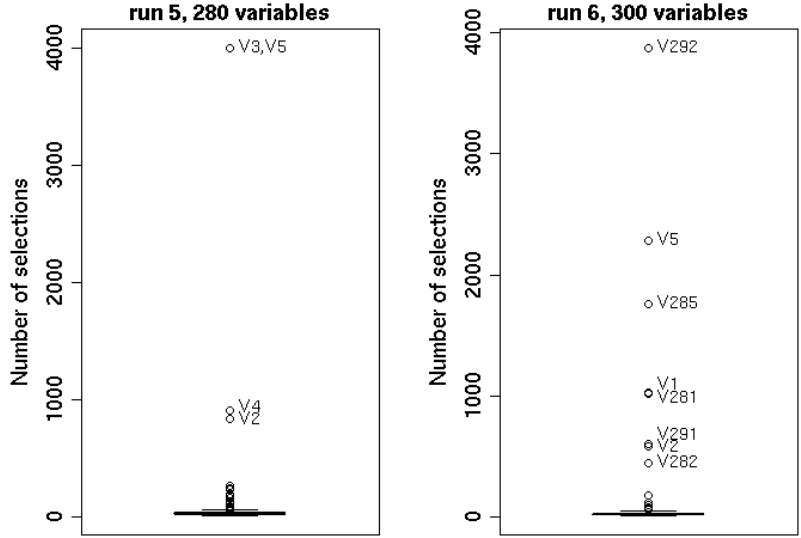}
			\end{center}
			\caption{\small{Boxplots of the number of selections of a variable after the burn-in period. Each point represents a variable (or several variables if superposed). The left boxplot corresponds to the run 5 with 280 variables. The right boxplot corresponds to the run 6 with 300 variables.}}
			\label{Boxplot:boxplotSimu}
		\end{figure}

		A final selection was performed for each of the 20 runs. Figure \ref{Boxplot:boxplotSimu} presents two  boxplots associated to   280 variables and  300 variables, respectively.  For the run with 280 variables there is a gap between the variables $V2,V3,V4$ and $V5$ and the others, hence we selected these four variables. For the run with 300 variables,  there is a gap between the variables selected in more than 400 iterations and the others, hence we selected the eight corresponding variables.

		Table \ref{tab10runs} gives the variables kept in the final selections of the 20 runs.
		Among the runs with the first 280 variables, 3 among the 5 variables used to generate the data were in the final selection of almost all runs, and the variables $V4$ was in the final selection of half of the runs. Notice that $V1$ was in none of the final selections. Among the runs with 300 variables, the variables $V1,V2,V3$ and $V5$ were present in most of the final selections, directly or indirectly through linear combinations. Contrarily to the runs with 280 variables, the variables $V4$ or $V284$ were in none of the final selections, while the variables $V1$ and $V281$ were in all the final selections. Concerning $V4$, it was indirectly in all the final selections through $V292$, which is a linear combination of $V3$ and $V4$. Eventually, the final selections of the runs with 300 variables appeared as relevant as the final selections of the runs with 280 variables, despite the fact that some variables were linear combinations of others.
		\begin{rmk}
		We obtained similar results with only 500 burn-in iterations and 500 post burn-in iterations, except that the variable $V4$ was in none of the final selections.
		\end{rmk}
		\begin{table}[h!]
			\begin{center}
			\begin{tabular}{|c|c|c|}
			\hline
			Variables & Number of selections & Number of selections\\
			 & among the 10 runs & among the 10 runs\\
			 & with 280 variables & with 300 variables\\
			\hline
			$V1$ & 0 &10\\
			$V2$ & 9 & 8\\
			$V3$ & 10 & 2\\
			$V4$ & 5 & 0\\
			$V5$ & 10 & 10\\
			\hline
			$V281=2 \times V1$ &  & 10\\
			$V282=2 \times V2$ &  & 9\\
			$V283=2 \times V3$ &  & 3\\
			$V284=2 \times V4$ &  & 0\\
			$V285=2 \times V5$ &  & 10\\
			$V291= V1 + V2$ &  & 7\\
			$V292= V3 - V4$ & \multirow{-5}{*}{Not available} & 10\\
			\hline
			\end{tabular}
			\caption{\small{Number of final selections among the 10 runs with the first 280 variables and among the 10 runs with 300 variables, for the variables $V1,V2,V3,V4,V5$ and linear combinations of these variables. No other variable was present in the final selections.}}
		\label{tab10runs}
		\end{center}
		\end{table}
		\FloatBarrier
		To assess the relevance of the final selections, predictions were performed.
		Concerning the runs with 300 variables, we can not fit a model with all the variables in final selections, because some of them are linear combinations of others. In that case we decided to fit a probit mixed model on the training set with the five linearly independent variables $V281,V282,V283, V285$ and $V292$. Sensitivity and specificity results are presented in Table \ref{tabsensispe}.
		For comparison, using the five variables used to generate the data, we obtained
		a sensitivity and  a specificity equal to 0.94 and 0.89. This is equivalent to results obtained using the selected variables $V281,V282,V283, V285$ and $V292$.

		\begin{table}[h!]
			\begin{center}
			\begin{tabular}{|c|c|c|c|c|c|}
			\hline
			\multicolumn{3}{|c|}{\cellcolor{lightgray}Variables selected among 280} & \multicolumn{3}{c|}{\cellcolor{lightgray}Variables selected among 300} \\
			\hline
			Variables & Sensitivity & Specificity & Variables & Sensitivity & Specificity\\
			\hline
			$V2, V3, V5$ & 0.87 & 0.89 & $V281$, $V282$ &  & \\
			 &  &  & $V283$, $V285$ & 0.94 & 0.89\\
			$V2, V3, V4, V5 $ & 0.93 & 0.96 & and $V292$ &  & \\
			\hline
			\end{tabular}
		\caption{\small{Sensitivity and specificity on the validation dataset.}}
		\label{tabsensispe}
		\end{center}
		\end{table}
		\FloatBarrier

\begin{figure}[h!]
			\begin{center}
			\includegraphics[scale=0.5]{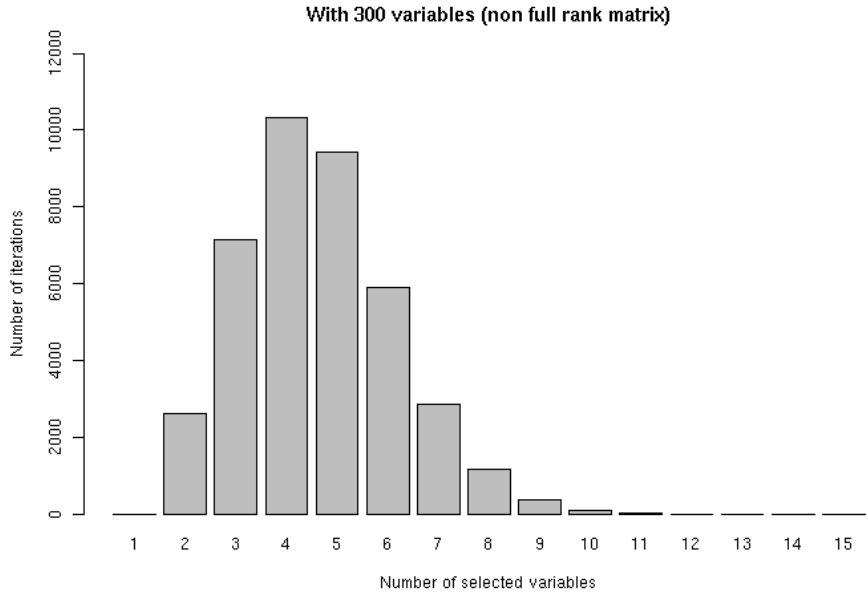}
			\end{center}
			\caption{\small{Number of iterations of the runs associated with a number of selected variables from 1 to 15. For the 10 runs, there were a total of 40000 post burn-in iterations.}}
			\label{Barplot:nbvarSimuSing}
		\end{figure}

		The number of components of $\gamma$ equal to 1  can vary from one iteration to another.
		Figure \ref{Barplot:nbvarSimuSing} shows, for the 10 runs with 300 variables, the number of iterations of the runs associated with a number of selected variables from 1 to 15. 
		Similar results were obtained for the 10 runs with the first 280 variables.

		\FloatBarrier

\begin{figure}[h!]
			\begin{center}
			\includegraphics[scale=0.45]{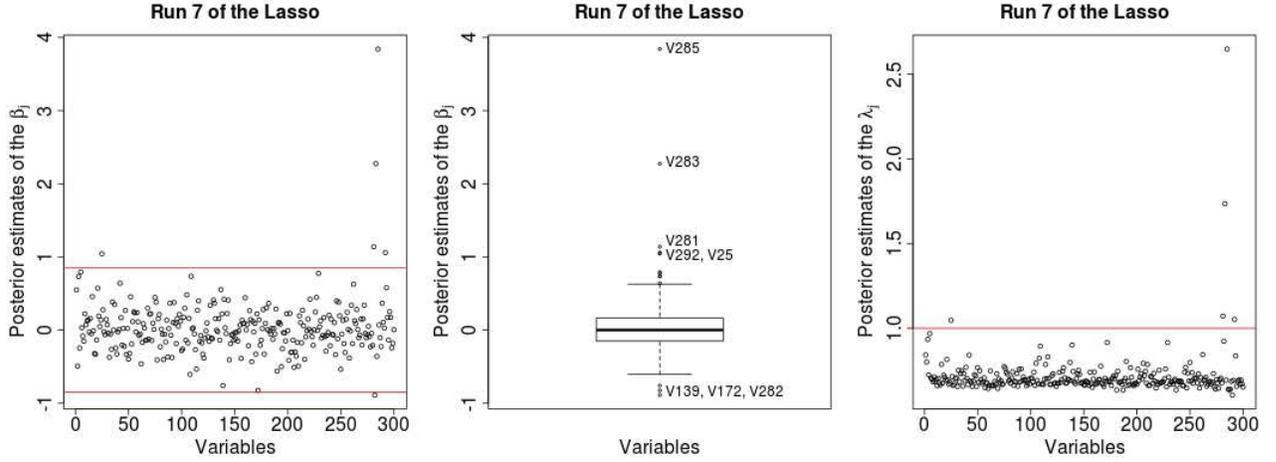}
			\end{center}
			\caption{\small{Results of the 7th run of the Bayesian Lasso. On the left are represented the values of the $\beta_j$s as well as the threshold used. Another way to represent these values is by using a boxplot, represented in the middle. On the right are represented the values of the $\lambda_j$s and the threshold used.}}
			\label{Lassosimu7}
		\end{figure}

	    \paragraph{Bayesian Lasso approach}~~\\
		Ten runs of the Bayesian Lasso were performed, with 5000 burn-in iterations and 15000 post-burn-in iterations. The results of the 7th run are illustrated in Figure \ref{Lassosimu7}.	
		\FloatBarrier
		 From this run, using the absolute values of the $\beta_j$s and a treshold of 0.85, the relevant variables $V281$, $V282$, $V283$, $V285$ and $V292$ were selected ($V284$ was indirectly selected by $V292$). However, we can note that the other relevant variables were not selected, and in particular the variables which are linear combinations of the selected variables.  Moreover, the non-relevant variable $V25$ was selected. Using the values of the $\lambda_j$s and a threshold of 1, the relevant variables $V281$, $V283$, $V285$ and $V292$ were selected, and the non-relevant variable $V25$ was still selected. Finally, using posterior CI, only the variables $V283$ and $V285$ were selected.\\
		 The variables selected during the ten runs are given in Table \ref{tablelassosimu}.
		\begin{table}[h!]
			\begin{center}
			\begin{tabular}{|c|c|c|c|}
			\hline
			\rule[-6pt]{0cm}{20pt} Variables & Using the $|\beta_j|$s & Using the $\lambda_j$s & Using the posterior CI\\
			\hline
			\rule[-6pt]{0cm}{20pt} selected in 9 runs & $V285$ & $V285$ & $V285$\\
			\hline
			\rule[-6pt]{0cm}{20pt} selected in 7 runs & $V283,V292$ & $V283$ & \\
			\hline
			\rule[-6pt]{0cm}{20pt} selected in 6 runs & & $V292$ & $V283$ \\
			\hline
			\rule[-6pt]{0cm}{20pt} selected in 3 runs & $V282$ & & \\
			\hline
			\rule[-6pt]{0cm}{20pt} selected in 2 runs & $V281$ & $V281,V282$ & \\
			\hline
			\rule[-6pt]{0cm}{20pt}  & $V25,V208,V209,V250$ & $V25,V250,V87,V127$ & $V250,V87,V127$\\
			\rule[-6pt]{0cm}{20pt}	\multirow{-2}{*}{selected in 1 runs}	   & $V87,V127,V270$ & $V270$ & \\
			\hline
			\end{tabular}
			\caption{\small{Variables selected during the 10 runs of the Bayesian Lasso approach, using three different methods, see \ref{lasso}.}}
		\label{tablelassosimu}
		\end{center}
		\end{table}
		\FloatBarrier

		Generally, it appeared that using the values of the $\beta_j$s and the $\lambda_j$s enabled us to select more relevant variables than using the posterior CI, but at the price of also selecting non-relevant variables. Besides, we can note that this Bayesian Lasso approach did not give very stable results. Indeed, some runs gave relevant selections of variables, like the run 1 for instance from which the variables $V281,V282,V283,V285$ and $V292$ were selected, while other runs gave selections of variables quite less relevant, like the run 10 from which the variables $V127$ and $V270$ were selected, or like the run 4 from which the variables $V285,V208,V209$ and $V250$ were selected. Between these two cases, some runs enabled us to select some relevant variables, but not all of them, like the runs 2, 5, 8 and 9 from which the variables $V283,V285$ and $V292$ were selected. Eventually, it appeared that the subsets of selected variables obtained were less stable than those obtained by the SSVS approach using the prior with a ridge parameter, and that more ``noise'' was observed since several non-relevant variables which were selected in only one of the ten runs.

	\subsection{Illustrations through real data}\label{realdata}
		As an illustration, Affymetrix microarray experiment results from patients with breast cancer were used. Data used in \cite{Baragatti1} were considered, see there for more details. Briefly, the patients come from three different hospitals, and the objective was to select some variables (probesets) which are indicative of the activity of the estrogen receptor (ER) gene in breast cancer. The hospital was considered as a random effect in the model, thus accounting for the different experimental conditions between the three hospitals. For each patient, the expressions of 275 probesets were kept, among which some were known to be relevant to explain the ER status (corresponding to variables 148, 260, 263 and 273). We used a training set made of 100 patients, and a validation set of 88 patients.
		In order to have a potentially singular $\mathbf{X}_{\gamma}^T\mathbf{X}_{\gamma}$ matrix, we added three variables to the data matrix $\mathbf{X}$. These variables were linear combinations of the known relevant variables, hence $\mathbf{X}$ was no more of full rank: $V276=2 \times V148$, $V277=-V260$ and  $V278=V263+V273$. We had only one random effect, which corresponded to the different hospitals. The hospitals are supposed independent, hence we put $D=\sigma^2 I_3$.

	    \paragraph{SSVS approach using the prior with a ridge parameter}~~\\
		We performed 10 runs of the sampler using only the first 275 variables, and 10 runs using all the 278 variables.
		In these two cases and for each run the same 100 patients and the same parameters were used.
		As in the previous illustration we chose $\tau_0=50$.
		The parameters $\lambda$ and $\tau$ were chosen as explained in \ref{calibrate}, yielding $\lambda=1/275$ and $c=50.0009$ when using 275 variables, and $\lambda=1/278$ and $c=50.00088$ when using 278 variables.

\begin{figure}[h!]
			\begin{center}
			\includegraphics[scale=0.5]{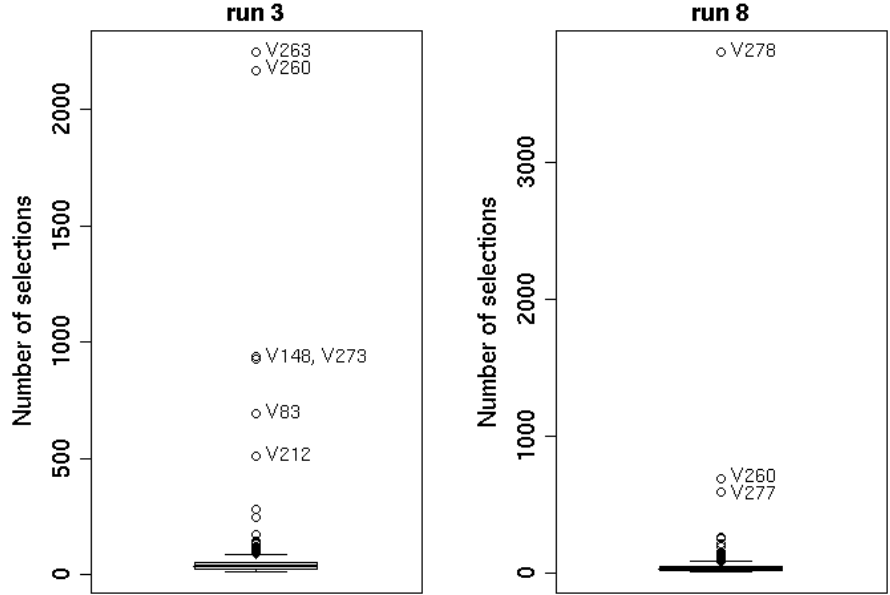}
			\end{center}
			\caption{\small{Boxplots of the number of selections of a variable after the burn-in period. The left boxplot corresponds to the run 3 with 275 variables and the right boxplot corresponds to the run 8 with 278 variables.}}
			\label{Boxplot:boxplotEx}
		\end{figure}

		Figure \ref{Boxplot:boxplotEx} presents a boxplot of a run with 275 variables and a boxplot of a run with 278 variables.
		Following the same reasoning as in the previous example, six
		variables were selected from the left boxplot and three  from the right boxplot.

		Table \ref{varkept} gives the variables kept in the final selections of the 20 runs. As in the previous example, the final selections of the runs with 278 variables appeared as relevant as the final selections of the runs with 275 variables, despite the fact that some variables were linear combinations of others.

		\begin{table}[h!]
		\vspace{0.5cm}
			\begin{center}
			\begin{tabular}{|c|c|c|c|}
			\hline
			Variables & Corresponding & Number of selections & Number of selections\\
			 & probesets & among the 10 runs & among the 10 runs\\
			 & & with 275 variables & with 278 variables\\
			\hline
			$V260$ & 228241\_at & 10 & 3\\
			$V273$ & 205862\_at & 9 & 0\\
			$V148$ & 209604\_s\_at & 5 & 1\\
			$V263$ & 228554\_at & 10 & 0\\
			$V83$ & 203628\_at & 7 & 0\\
			$V66$ & 202088\_at & 1 & 0\\
			$V212$ & 215157\_x\_at & 1 & 0\\
			\hline
			$V277=-V260$ & collinearity &  & 3\\
			$V278=V263+V273$ & linear combination & \multirow{-2}{*}{Not available}  & 10\\
			\hline
			\end{tabular}
			\caption{\small{Number of final selections among the 10 runs with the first 275 variables and among the 10 runs with 278 variables, for the different variables and linear combinations. No other variable was present in the final selections.}}
			\label{varkept}
			\end{center}
		\end{table}
		\FloatBarrier

		Predictions were also performed. Table \ref{tabSS2} contains sensitivity and specificity results. For comparison, using the four relevant variables $V260,V263,V148$ and $V273$, we obtained a sensitivity equal to 0.94 and a specificity equal to 1. This is equivalent to results obtained using only the two variables  $V278$ and $V277$.

		\begin{table}[h!]
			\begin{center}
			\begin{tabular}{|c|c|c|c|c|c|}
			\hline
			\multicolumn{3}{|c|}{\cellcolor{lightgray}Variables selected among 275} & \multicolumn{3}{c|}{\cellcolor{lightgray}Variables selected among 278} \\
			\hline
			Variables & Sensitivity & Specificity & Variables & Sensitivity & Specificity\\
			\hline
			$V260, V273, V263$ & 0.92 & 1 & $V278$ & 0.87 & 0.97\\
			 & &  & $V278$, $V277$ & 0.94 & 1\\
			\hline
			\end{tabular}
		\caption{\small{Sensitivity and specificity on the validation dataset.}}
		\label{tabSS2}
		\end{center}
		\end{table}
		\FloatBarrier

\begin{figure}[h!]
			\begin{center}
			\includegraphics[scale=0.45]{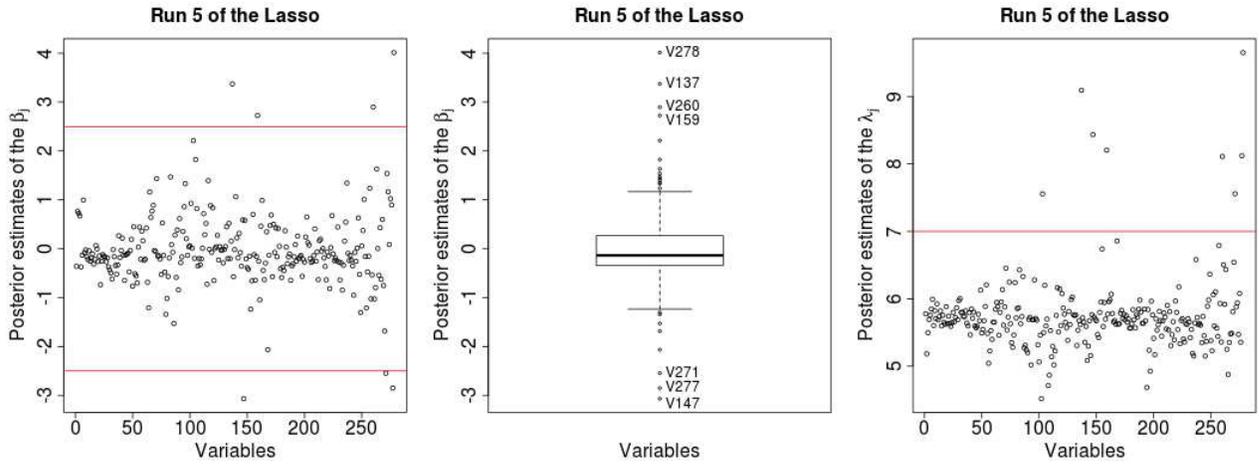}
			\end{center}
			\caption{\small{Results of the 5th run of the Bayesian Lasso. On the left are represented the values of the $\beta_j$s as well as the threshold used. Another way to represent these values is by using a boxplot, represented in the middle. On the right are represented the values of the $\lambda_j$s and the threshold used.}}
			\label{LassoExample5}
		\end{figure}

	    \paragraph{Bayesian Lasso approach}~~\\
		Ten runs of the Bayesian Lasso were performed, with 5000 burn-in iterations and 15000 post-burn-in iterations. The results of the 5th run are illustrated in Figure \ref{LassoExample5}.	
		\FloatBarrier
		 From this run, using the absolute values of the $\beta_j$s and a treshold of 2.5, the relevant variables $V260$, $V278$ and $V277$ were selected. Moreover, the less relevant variable $V137$, $V159$, $V147$, $V271$ were selected. Using the values of the $\lambda_j$s and a threshold of 7, the relevant variables $V260$, $V278$ and $V277$  were selected, and the less relevant variables $V103$, $V137$, $V159$, $V147$ and $V271$ were selected. Finally, using posterior CI, no variable was selected.\\
		 The variables selected during the ten runs are given in Table \ref{tablelassoexample}.
		\begin{table}[h!]
			\begin{center}
			\begin{tabular}{|c|c|c|c|}
			\hline
			\rule[-6pt]{0cm}{20pt} Variables & Using the $|\beta_j|$s & Using the $\lambda_j$s & Using the posterior CI\\
			\hline
			\rule[-6pt]{0cm}{20pt} selected in 7 runs & $V278$ & $V278$ & \\
			\hline
			\rule[-6pt]{0cm}{20pt} selected in 4 runs & $V277,V260$ &  &  \\
			\hline
			\rule[-6pt]{0cm}{20pt} selected in 3 runs &  & $V260,V263,V277$ & \\
			\hline
			\rule[-6pt]{0cm}{20pt}  & $V263,V114,V140,V272$ & $V140,V145,V83$ & $V278,V145,V271,V266$\\
			\rule[-6pt]{0cm}{20pt} \multirow{-2}{*}{selected in 2 runs} & $V147,V271,V83$ & & \\
			\hline
			\rule[-6pt]{0cm}{20pt}  & $V276,V80,V145,V71$ & $V114,V80,V252,V71$ & $V114,V140,V272,V147$\\
			\rule[-6pt]{0cm}{20pt}		   & $V102,V137,V159,V59$ & $V78,V215,V165,V102$ & $V84,V83,V95,V115$\\
			\rule[-6pt]{0cm}{20pt}		   & $V84,V105,V78,V95$ & $V60,V137,V159,V147$ & $V161,V105,V272,V276$\\
			\rule[-6pt]{0cm}{20pt}		   & $V161,V266,V105,V2$ & $V271,V103,V59,V84$ & \\
			\rule[-6pt]{0cm}{20pt}	\multirow{-6}{*}{selected in 1 runs}	   & $V161,V266,V105,V2$ & $V106,V2,V105,V266,V161$ & \\
			\hline
			\end{tabular}
			\caption{\small{Variables selected during the 10 runs of the Bayesian Lasso approach, using three different methods, see \ref{lasso}.}}
		\label{tablelassoexample}
		\end{center}
		\end{table}
		\FloatBarrier

		The same general remarks than those made from the simulated example can be done (see \ref{simulateddata})

		\begin{rmk}
		 We compared the SSVS approach using the prior with a ridge parameter with the Bayesian Lasso. Note that in the case of smaller problems, the marginal likelihoods of all the models can be calculated using the method of \cite{Chib1995}, and it is then possible to confront them to the results obtained by the SSVS or the Bayesian Lasso approaches.
		\end{rmk}

	\subsection{Sensitivity analysis}\label{sensitivity}
		Concerning the variable selection coefficient $\tau$, the method of variable selection without the ridge parameter is not sensitive to its value (see \cite{Baragatti1}), but it is mainly due to the fact that the number of variables selected at each iteration of this algorithm was fixed. It is no more the case for the algorithm proposed in this paper, hence it seems necessary to assess its sensitivity to this parameter. 
		Therefore we studied the influences of $\tau$  and $\lambda$ when they are chosen as proposed in Section \ref{calibrate}, or arbitrarily.
		We also looked at the behavior of the algorithm when the value of the $\pi_j$, the prior distribution parameters of $\sigma^2$ and the number of iterations vary.\\
		For this sensitivity study we used the example with simulated data (Section \ref{simulateddata}) with 300 variables. The different values of the parameters are presented in Table \ref{Tab:tabsensiridge}. In this table, the number of relevant variables in the final selections of the runs are given, the twelve relevant variables being $V1,V2,V3,V4,V5,V281$, $V282,V283,V284,V285,V291$ and $V292$. The sensitivity was assessed by using the relative weighted consistency measure of \cite{Somol2008}, denoted by $CW_{rel}$. It is a measure evaluating how much subsets of selected variables for several runs overlap, and it shows the relative amount of randomness inherent in the concrete variable selection process. It takes values between 0 and 1, where 0 represents the outcome of completely random occurrence of variables in the selected subsets and 1 indicates the most stable variable selection outcome possible.

		\begin{table}[!h]
			\hspace{-1cm}
			\begin{tabular}{|c|c|c|c|c|c|c|c|c|}
			\hline
			\rowcolor{lightgray}  &  & &  & Value & Prior & Iterations & Nb of & \\
			\rowcolor{lightgray} Run & $\tau_0$ & $\tau$ & $\lambda$ & of $\pi_j$ & for & post burn-in & relevant & $\mathcal{S}$\\
			\rowcolor{lightgray}  &  & &  & $\forall j$ & $\sigma^2$ & (burn-in) & variables & \\
			\hline
			1 & 10 & 10.00035 (\ref{choixtau}) &  &  &  &  & 3  & \\
			2 & 50 & 50.00885 (\ref{choixtau}) &  &  &  &  &  8 & \\
			3 & 100 & 100.0354 (\ref{choixtau}) &  &  &  &  &  8 & \\
			4 & 1000 & 1003.553 (\ref{choixtau}) &  &  &  &  &  8 & \\
			5 & 10000 & 10367.03 (\ref{choixtau}) & \multirow{-5}{*}{$1/p=1/300$} & \multirow{-5}{*}{5/300} & \multirow{-5}{*}{$\mathcal{IG}(1,1)$} & \multirow{-5}{*}{4000 (1000)} & 8  & \multirow{-5}{*}{0.857}\\
			\hline
			6 &  &  & $1/p=1/300$ &  &  &  &  8 &  \\
			7 &  &  & $100/p=1/3$ &   &  &  &  8 & \\
			8 &  \multirow{-3}{*}{(\ref{choixtau}) non} &  & $1$ &  &  &  & 8  &  \\
			9 &  \multirow{2}{*}{used} &  & $10$ &   &  &  & 8 &  \\
			10 & & \multirow{-5}{*}{100} & $100$ & \multirow{-3}{*}{5/300}  & \multirow{-3}{*}{$\mathcal{IG}(1,1)$} & \multirow{-3}{*}{4000 (1000)} & 3  & \multirow{-5}{*}{0.8}\\
			\hline
			11 &  & 10 & $1/p=1/300$ &  &  &  & 5 &   \\
			12 &  & 10 & 10 &  &  &  & 3  &  \multirow{-2}{*}{0.348}\\
			13 & \multirow{-3}{*}{(\ref{choixtau}) non} & 1000 & $1/p=1/300$ &   &  &  & 0  & (0.639\\
			14 & \multirow{2}{*}{used}  & 1000 & 10 &  &  & &  8 & without \\
			15 & & 100 & $100/p=1/3$ & \multirow{-5}{*}{5/300} & \multirow{-5}{*}{$\mathcal{IG}(1,1)$} & \multirow{-5}{*}{4000 (1000)}  & 8  & run 13) \\
			\hline
			16 &  &  &  & $5/300$ &  &  & 8 &  \\
			17 &  &  &  & $50/300$ &   &  & 12 &  \\
			18 & \multirow{-3}{*}{100} & \multirow{-3}{*}{100.0354 (\ref{choixtau})} & \multirow{-3}{*}{1/p=1/300} & $100/300$ & \multirow{-3}{*}{$\mathcal{IG}(1,1)$} &  \multirow{-3}{*}{4000 (1000)} &  12 & \multirow{-3}{*}{0.848}\\
			\hline
			19 &  &  &  &  & $\mathcal{IG}(1,1)$ &  & 8 &  \\
			20 &  &  &  &  & $\mathcal{IG}(2,5)$ &  &  8 & \\
			21 & \multirow{-3}{*}{100} & \multirow{-3}{*}{100.0354 (\ref{choixtau})} & \multirow{-3}{*}{1/p=1/300} & \multirow{-3}{*}{5/300} & $\mathcal{IG}(5,2)$ & \multirow{-3}{*}{4000 (1000)} & 8 & \multirow{-3}{*}{1} \\
			\hline
			22 &  &  &  &  &  & 500 (500) & 8 &  \\
			23 &  &  &  &  &   & 4000 (1000) & 8  & \\
			24 & \multirow{-3}{*}{100} & \multirow{-3}{*}{100.0354 (\ref{choixtau})} & \multirow{-3}{*}{1/p=1/300} & \multirow{-3}{*}{5/300}  & \multirow{-3}{*}{$\mathcal{IG}(1,1)$} & 40000 (10000) & 8  & \multirow{-3}{*}{1}\\
			\hline
			\end{tabular}
			\caption{\small{Parameters of the runs for the sensitivity study and associated relative weighted consistency measure of Somol and Novovicova $CW_{rel}$. 
			}}
			\label{Tab:tabsensiridge}
		\end{table}

\begin{figure}[!h]
			\begin{center}
			\includegraphics[scale=0.5]{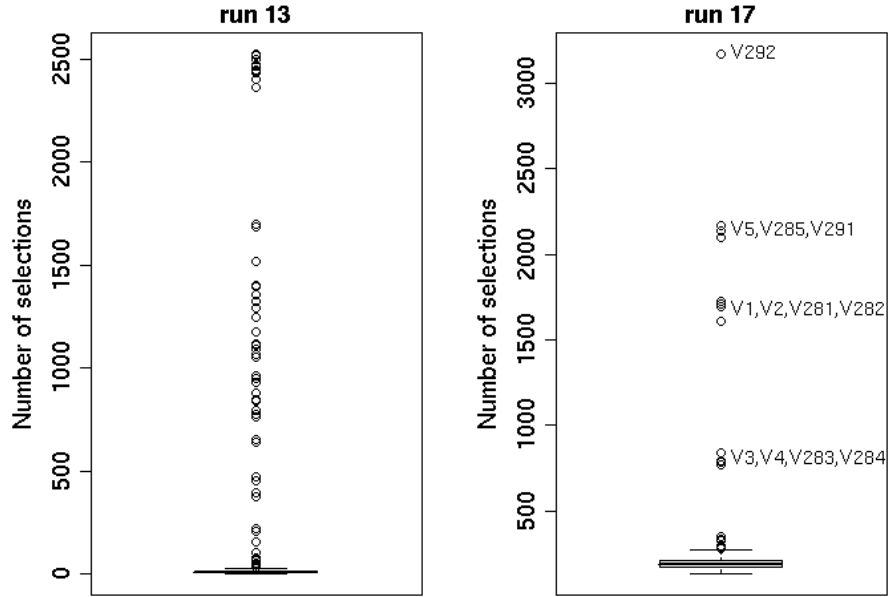}
			\end{center}
			\caption{\small{Boxplot of the number of selections of a variable after the burn-in period, for two runs with 300 variables.}}
			\label{Boxplot:boxplotsensibSimu}
		\end{figure}

\begin{figure}[!h]
			\begin{center}
			\includegraphics[scale=0.4]{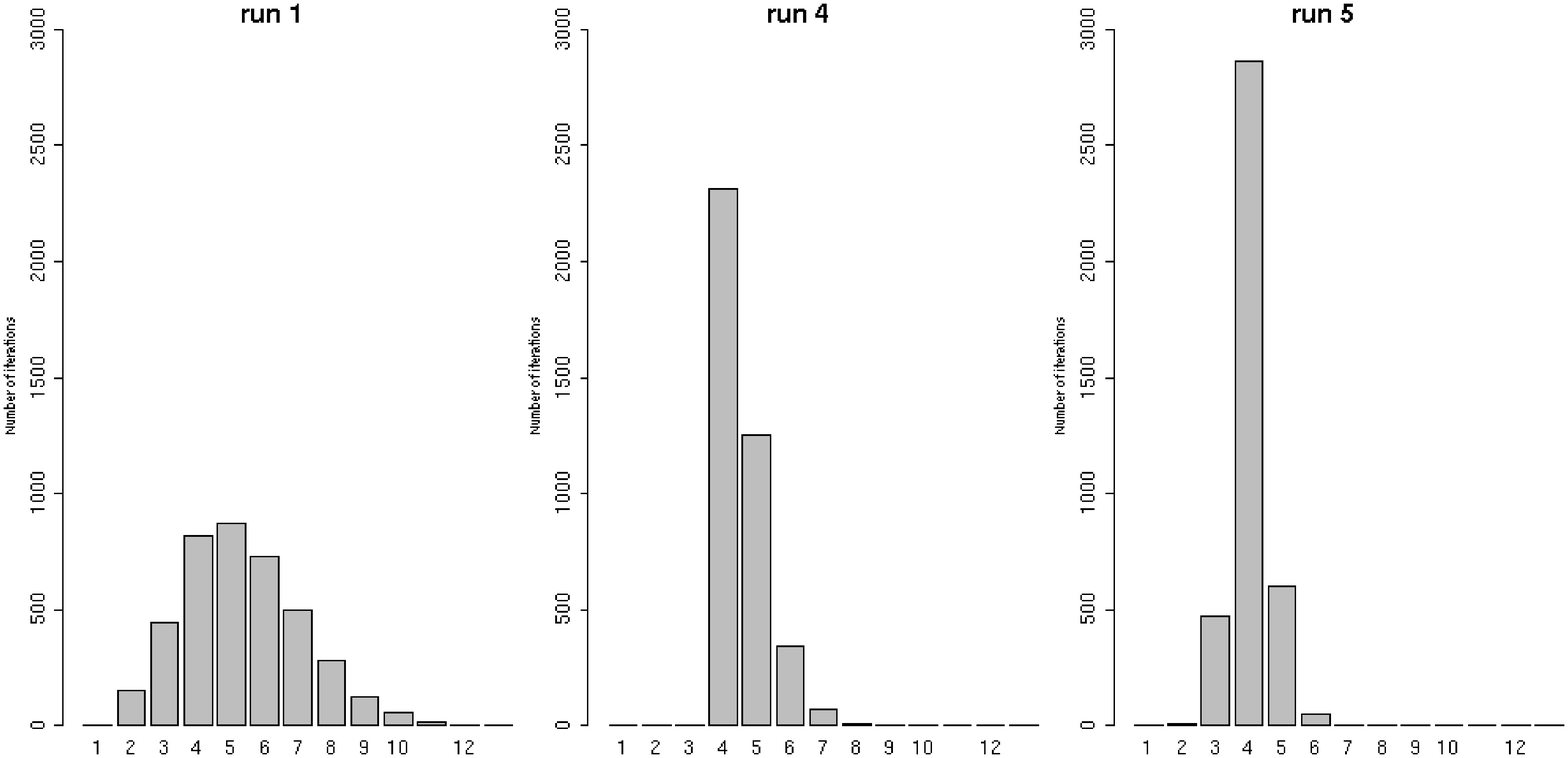}
			\end{center}
			\caption{\small{Number of iterations of the runs 1,4 and 5 associated with a number of selected variables from 1 to 14. For each run, there were 4000 post burn-in iterations.}}
			\label{nbvarSimuNonSing-c}
		\end{figure}

\begin{figure}[!h]
			\begin{center}
			\includegraphics[scale=0.4]{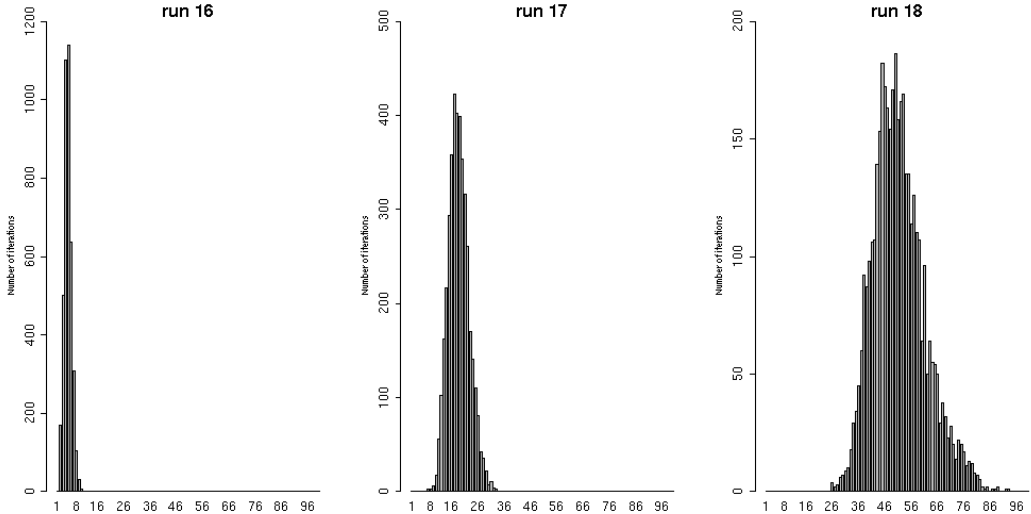}
			\end{center}
			\caption{\small{Number of iterations of the runs 16,17 and 18 associated with a number of selected variables from 1 to 100. For each run, there were 4000 post burn-in iterations.}}
			\label{nbvarSimuNonSing-pi}
		\end{figure}

		The algorithm was generally not sensitive to the values of the hyper-parameters, since most of the relevant variables were usually  selected.
		The boxplots obtained were often similar to the right boxplot of Figure \ref{Boxplot:boxplotSimu}. In particular, the algorithm was not overly sensitive to the values of $\tau$ and $\lambda$. There was only one run (the 13th) where no variable could be really distinguished from others, and none of the top-ranked variables was a relevant one, see Figure \ref{Boxplot:boxplotsensibSimu}.
		This run corresponds to a large $\tau$ and a small $\lambda$.
		The runs 17 and 18 are also noticeable, as all relevant variables were finally selected, see Figure \ref{Boxplot:boxplotsensibSimu}. They correspond to high values of $\pi_j$, and the cost for these relevant runs was longer computational times. Eventually, we observed that the values of $\tau$ and $\pi_j$ play a role in the number of variables selected at each iteration of the algorithm. The value of $\tau$ modified the distribution of this number, see Figure \ref{nbvarSimuNonSing-c}. Besides, this number increased with the value of $\pi_j$, see Figure \ref{nbvarSimuNonSing-pi}. However, even if the number of variables selected at each iteration of the algorithm was high, it did not influence the final selections of the runs, and it did not influence the number of variables which were distinguishable from others.

		\FloatBarrier

\section{Discussion}\label{Discussion}
	Classical stochastic search variable selection methods often propose the use of the $g$-prior of Zellner. This prior can not be used if $p>n$, or if some variables are linear combinations of others. In particular, this last case can occur when several datasets with common covariates are merged. The prior for $\beta_{\gamma}$ studied in this manuscript is a possible alternative, and is a reparametrization  of the prior of \cite{GuptaIbrahim} where $\tau$ and $\lambda$ can be chosen independently. In this case the parameter $\tau$ does not influence the coefficient of the identity matrix. Using this prior, a way to jointly choose $\tau$ and $\lambda$ was suggested and the results obtained on simulated data and on a real dataset were good and stable, whether some variables were linear combinations of others or not. Moreover, when $\tau$ and $\lambda$ were chosen independently, the proposed method proved to be robust to the choices of these hyper-parameters, as shown in the sensitivity analysis.\\
	
	In practice, even if $\mathbf{X}_{\gamma}^T \mathbf{X}_{\gamma}$ is theoretically invertible, some variables can be highly correlated and $\mathbf{X}_{\gamma}^T \mathbf{X}_{\gamma}$ can be computationally singular. Moreover, we do not necessarily know if some variables are linear combinations of others and to avoid a computational problem  we suggest in any cases to use the prior and the algorithm proposed in this paper.
	Once a final selection of variables benoted by $\gamma +$ is obtained by our algorithm, the rank of the matrix with all the variables finally selected, denoted by $X_{\gamma +}$, should be computed. If this matrix is not of full rank,  we can take a submatrix of $X_{\gamma +}$ of full rank as a new data matrix. Note that it is easier to take linearly independent columns of $X_{\gamma +}$, than linearly independent columns of $X$, especially if $p$ is quite large.\\

	We compared the results of the proposed SSVS method using a prior with a ridge parameter, with results obtained from the competing approach of the Bayesian Lasso. This last approach can also be used when some variables are linear combinations of others or if $p>n$. Moreover, its implementation is quite easy, and does not necessitate Metropolis-Hastings steps. Compared to the SSVS method proposed in \ref{SSVS}, an iteration is then less computing demanding. 
	However, the Bayesian Lasso approach seems less practical than the SSVS approach. Indeed, in formula (\ref{fullbetalasso}) it is $\mathbf{X}$ which is used and not a submatrix $\mathbf{X}_{\gamma}$ like in the SSVS approach. The computing time to multiply or to invert matrices is then higher, and it could become an issue if $p$ is very large. Besides, on the previous simulation and example, it appeared than more iterations are needed by the Bayesian Lasso compared to the SSVS approach (20000 vs 5000). 
	Concerning the results, it appeared that the runs of the Bayesian Lasso were less stable than those of the SSVS approach, and that more ``noise'' was observed, since several non or less relevant variables were selected in only one of the ten runs. 
	It is important to note that we adapted a simple Bayesian Lasso approach to probit mixed models, but many extensions of the classical Lasso exist and can be adapted in Bayesian approaches, like the fused Lasso (\cite{Tibshirani2005}), the group Lasso (\cite{YuanLin2006}) or the Elastic Net (\cite{ZouHastie2005}) for instance. Recently, \cite{Kyung2010} showed how to adapt this extensions in Bayesian approaches, and it could be interesting to compare them to the SSVS approach.\\

	Several extensions of the proposed SSVS method using a prior with a ridge parameter can be done in the future.
	First, in classical cases using the $g$-prior, many authors suggested to put prior distributions on $\tau$, see Section \ref{intro}. Following them, an idea could be to put prior distributions on the hyper-parameters $\tau$ and $\lambda$. However, these authors often used Bayes Factors \citep[see for instance][]{CeleuxMarinRobert} and not a latent $\gamma$ vector as done in this paper. They were then more in the spirit of model selection than in the spirit of variable selection. 
	Finally, it would be interesting to have a non-supervised criterion to decide which variables should be in the final selection of a run. We suggested to represent the vector of the numbers of iterations during which variables have been selected by a boxplot, and to use a treshold to decide which variables should be in the final selection. However, we could develop a more formal criterion.

\bibliography{references}

\end{document}